\documentclass[aps,prd,reprint,superscriptaddress,nofootinbib]{revtex4-2} 

\pdfoutput=1
\usepackage{comment}
\usepackage{tablefootnote}
\usepackage{graphicx}
\usepackage[T1]{fontenc}
\usepackage{lmodern} 
\usepackage{microtype}
\usepackage[caption=false]{subfig} 
\usepackage{bm}
\usepackage{tensor}
\usepackage{amsmath, amssymb}
\usepackage{dcolumn}
\usepackage{cancel}
\usepackage{booktabs}
\usepackage{multirow}
\usepackage[usenames,dvipsnames]{color}
\usepackage{array}
\usepackage{orcidlink}

\hypersetup{
     breaklinks=true,
    pdfstartview={FitH},    
    colorlinks=true,       
    linkcolor=blue,          
    citecolor=blue,        
    filecolor=red,      
    urlcolor=magenta,           
    anchorcolor=green,      
    linktocpage=true
}
\usepackage{cleveref}

\def\doi{http://doi.org}

\newcommand{\HCd}{\mathcal{H}}

\def\HCdt0{\tilde{\HCd}_{0}}

\newcommand{\afftech}{The Technion Department of Physics, The Technion – Israel Institute of Technology, Haifa 3200003, Israel}

\begin{document}
\title{Probing the Dark Sector using the Fornax Satellite}

\author{E. Zamlung\,\orcidlink{0000-0001-6623-6014}}
\email{eyalzamlung@campus.technion.ac.il}
\affiliation{\afftech}

\author{A. Nusser\,\orcidlink{0000-0002-8272-4779}}
\email{adi@physics.technion.ac.il}
\affiliation{\afftech}


\date{\today} 

\begin{abstract}

A long-range force acting only in the dark sector can displace the stellar component of satellite galaxies relative to their dark matter (DM) halos, thereby \emph{breaking} the weak equivalence principle (WEP) between DM and baryons.
We investigate observational signatures of such WEP breaking using $N$-body simulations of a Fornax-like Milky Way (MW) satellite, implementing a fifth force between DM particles of amplitude $\beta$ relative to Newtonian gravity (with a screening length much larger than the MW halo size).
We find that $\beta \gtrsim 0.6$ strips and unbinds the bulk of the stellar component, leaving at most a negligible bound remnant that fails to match the observed stellar content, surface-brightness, and line-of-sight velocity-dispersion profiles of Fornax.
By contrast, $\beta \lesssim 0.2$ yields only mild stellar stripping and remains broadly consistent with current photometric and kinematic constraints within our modeling assumptions.
\end{abstract}

\maketitle

\section{Introduction}

The $\Lambda$CDM model is the standard cosmological framework describing the large-scale structure and evolution of the universe.
It combines a cosmological constant ($\Lambda$) with cold dark matter (CDM) to account for cosmic expansion and structure formation \cite{PeeblesRevModPhys.75.559, Riess1998AJ, Planck2020AA}.
The model successfully explains key observations, including the anisotropies and polarization of the cosmic microwave background, and galaxy clustering statistics—assuming mild biasing between mass and galaxy distributions.

Despite its success, the DM  particle remains undetected apart from its gravitational signatures. Candidate particles include WIMPs, axion-like particles, sterile neutrinos, and primordial black holes \citep[e.g.,][]{CarrPBH,essig2013darksectorsnewlight}.
In addition, DESI measurements of the Baryon Acoustic Oscillation scale suggest a possible deviation from a pure cosmological constant, hinting at evolving dark energy \citep{DESI2025}.

The mere existence of DM  motivates extensions beyond the Standard Model of particle physics, including the possibility of long-range forces acting exclusively within the dark sector \cite{essig2013darksectorsnewlight}.
Given the complexity of interactions in the visible sector, it is natural to expect that the dark sector could exhibit similarly rich dynamics.
Proposed scenarios include dark photons mediating interactions within a hidden sector, scalar fields from modified gravity or string-inspired frameworks, and vector or tensor fields associated with Lorentz invariance violation \citep[e.g.,][]{Colladay1998LV, Jacobson2001}.
Such interactions could significantly alter the dynamics of DM  halos and structure formation, leaving distinct observational signatures.

Here, we are concerned with the implications of breaking the Weak Equivalence Principle (WEP) in the dark sector. Certain models introduce long-range forces acting exclusively between DM particles, leading to significant WEP breaking that can impact structure formation and potentially alleviate small-scale issues in $\Lambda$CDM  \cite{Keselman_2009} without invoking complex baryonic physics. 
Implications of such forces on structure formation have been explored in the literature with interesting effects \cite{Frieman:1993fv, Hellwing_2010, Hellwing_2013, Nusser_2005}.

 Our focus is on mass segregation between the stellar and DM components in satellite galaxies of the MW, induced by a fifth force that breaks the WEP.
This has been studied in detail for the Sagittarius (Sgr) dwarf galaxy and its stellar stream, composed of stars tidally stripped by the Galactic potential \cite{Frieman:1993fv, Kesden_2006oct, Kesden_2009}.

In the absence of WEP breaking, the gravitational tidal field is approximately symmetric in a frame centered on the satellite, as obtained by expanding the MW potential near the satellite's center.
This symmetry reflects the fact that gravity acts equally on stars and DM.

A fifth force acting only on the dark sector introduces a relative acceleration between the components, breaking this symmetry.
This effect was employed by \cite{Kesden_2006oct} to constrain the strength of the fifth force.

However, \cite{Keselman_2009} demonstrated that for a fifth force with strength comparable to gravity, the stellar component of Sgr would be abruptly yanked from its DM halo and continue evolving under the MW gravitational field alone.
In this scenario, the resulting stellar stream remains symmetric and thus consistent with observations \cite{Keselman_2009}.
These authors proposed that Sgr contains a gravitationally self-bound stellar core, free of DM, which survives the separation and is consistent with both the morphology of the observed stream and internal kinematics of Sgr.
In this model, the elevated stellar velocity dispersion—potentially exceeding the virial value—is attributed to Sgr's current proximity to pericenter in its Galactic orbit.

In the present paper, we extend this line of study to investigate the effects of a fifth force on the well-studied Fornax satellite galaxy of the MW.
The Fornax dwarf spheroidal galaxy (dSph) is located approximately $147 \mathrm{~kpc}$ from the Galactic Center, significantly farther than Sgr, and follows a much less eccentric orbit.
It is an elliptical galaxy with a major axis of about $710\, \mathrm{~pc}$ and a total mass of  $\approx {10^8} \, \mathrm{M}_{\odot}$, predominantly composed of DM \cite{Mateo_1998}. Another important distinction from Sgr is that the stellar component of Fornax is not gravitationally self-bound, and its low orbital eccentricity implies that its elevated velocity dispersion cannot be attributed to amplification by the Galactic tidal field.

We adopt the following potential for   the  fifth  force acting between two DM particles,
\begin{equation}
    \varphi(r)=-\frac{\beta G m^2}{r} e^{-r / r_{s}}\; ,
\label{eq:LRSI-potential}
\end{equation}
where $r$ is the separation, $m$ is the mass of each particle  and $\beta$ is a positive parameter dictating the force amplitude,  and  $r_{s}$ is the screening length  on the order of $1 \textrm{~Mpc}$ today, remaining constant in comoving coordinates.

The force law in \cref{eq:LRSI-potential} is partly motivated by string theory, where the DM has a long-range scalar force screened by light particles  \cite[e.g.][]{PhysRevD.70.123510}.
To avoid conflicts with laboratory tests of Newton's law, the scalar's coupling to the visible sector must be minimal, so the proposed modification in \cref{eq:LRSI-potential} applies solely to DM particles.

We shall assume that $r_\textrm{s}$ is larger than the size of the system, such that the fifth force is 
\begin{equation}
\label{eq:fifthgrav}
    \mathbf{F}_{5}= \beta \mathbf{F}_\textrm{gravity}\; .
\end{equation}
Our study will rely on several $N$-body simulations of the evolution of a Fornax-like system, taking into account as many observational constraints. These constraints include phase space information obtained by Gaia \cite{Fritz_2018} and the internal structure of Fornax \cite{Walker_2011, Amorisco_2013}. We also contrast the result with a simplified analytic model for tidal stripping in the presence of a fifth force of the form \cref{eq:fifthgrav}.

The outline of the paper is as follows. 
In \cref{sec:numerical}, we describe our numerical modeling, including the construction of the MW mass components, the Fornax dSph initial conditions, orbital setup, and the modifications introduced to the $N$-body code. 
In \cref{sec:results}, we present the $N$-body simulation results, highlighting the dependence on the fifth-force parameter $\beta$ and comparing the outcomes with observed properties of Fornax. 
In \cref{sec:conclusion}, we discuss the implications of our findings for long-range dark-sector forces and summarize the main conclusions. 
Finally, in the Appendices, we develop simplified analytic models: \cref{sec:analyticpotential} introduces the effective potential in the presence of a fifth force, followed by an approximation for stellar tidal stripping and an estimate of the critical value of $\beta$ for complete stellar–DM segregation. 

\section{Numerical Modeling}
\label{sec:numerical}

Fornax is represented as a two-component system of DM and stars. The MW model comprises three components: a baryonic disk (stars and gas), a stellar bulge, and a DM halo. The gravitational fields of the disk and bulge are held static. At the same time, the DM halo is represented by live simulation particles that evolve under their self-gravity, the fifth force, and the interaction forces with Fornax.

Both the stellar and DM components of Fornax are sampled by particles: stellar particles evolve under their self-gravity and the MW gravitational field, whereas DM particles additionally experience the fifth force arising from all DM particles in Fornax and in the MW halo.
 
There are two reasons for simulating the DM of the MW by live particles rather than modeling it as a predefined static gravitational potential. 
Firstly, the distribution of DM mass (after stabilization) is directly affected by the value of $\beta$ in the simulation, which directly influences the potential felt by the satellite along its entire orbit.
Secondly,  simulating DM halos is necessary to include dynamical friction on the 
 satellite, as it perturbs the mass distribution of the halo. 
 
\subsection{The MW Galaxy Potential}

\subsubsection{Static Potential Components}
A baryonic Miyamoto-Nagai \cite{Miyamoto_1975} gravitational  potential for the  MW disk is adopted,
 \begin{equation}
     \Phi_{\textrm{disk}}=-\frac{G M_{\textrm {d }}}{\sqrt{r_{\perp}^{2}+\left(a_{\textrm{d}}+\sqrt{z^{2}+d^{2}}\right)^{2}}}\, ,
     \label{eq:disk_potential}
 \end{equation}
 where $z$ and $r_{\perp}$ are cylindrical coordinates, with $z$ perpendicular to the disk.

The stellar  bulge potential is taken as  a spherical Hernquist  \cite{Hernquist_1990} in the form of,
\begin{equation}
    \Phi_{\textrm {bulge }}=-\frac{G M_{\textrm {b }}}{r+c}\, .
    \label{eq:bulge_potential}
\end{equation}
The parameters  $a_d$, $d$ and $c$  are given in \cref{tab:mw_parameters}
\subsubsection{The MW  Halo}
\label{subsection:MW}
The MW halo is represented  as  an equilibrium $N$-body  realization of a Dehnen density profile \cite{Dehnen_1993},
\begin{equation}
    {\rho_{\textrm{halo}}}(r)=\frac{(3-\gamma) {M_{\textrm{DM}}}}{4 \pi} \frac{a}{r^\gamma(r+a)^{4-\gamma}}\; .
\label{eq:dehnen_density}
\end{equation}

Both the scale radius $a$ and $\gamma$, which control the slope of the density profile near the center, are given as the best-fit values after the relaxation (see \cref{tab:mw_parameters}).

The initial conditions were generated with the \texttt{galstep} code \cite{galstep}, a versatile tool compatible with \texttt{GADGET2}, which we employ for our simulations.  \texttt{galstep} assigns particles statistically isotropic velocity distributions designed to approximate dynamical equilibrium.  After that, we evolve each profile in \texttt{GADGET2} through a relaxation run over roughly $\sim 10\times\textrm{ dynamical time}$ to ensure the system reaches a truly self-stable configuration.

\begin{table*}[ht]
\centering
\scalebox{0.93}{
    \begin{tabular}{@{}ccccc@{}}
    \toprule
    \textbf{Component} & \textbf{Modeling form} & \multicolumn{3}{c}{\textbf{Parameters}} \\ \midrule
    Disk    & Static potential  & ${M_\textrm{d}} = {10^{11}}{M_ \odot }$                  & ${a_{\textrm{d}}} = 6.5{\mathrm{~kpc}}$      & $d = 0.3{\mathrm{~kpc}}$ \\
    Bulge   & Static potential  & ${M_\textrm{b}} = 3.4 \times {10^{10}}{M_ \odot }$       & $c = 0.7{\mathrm{~kpc}}$      &                         \\
    DM halo & Simulated halo          & ${M_\textrm{DM}} = 1.2 \times {10^{12}}{M_ \odot }$ & ${a_{\textrm{initial}}} = 20{\mathrm{~kpc}}$ &   ${\gamma_{\textrm{initial}}} = 1$                      \\ \bottomrule
    \end{tabular}
}
\caption{Parameters for the MW model used in this study \cite{Keselman_2009,McMillan_2011,Schönrich_2010}.}
\label{tab:mw_parameters}
\end{table*}

\subsection{The Fornax dSph}

The initial DM and stellar distributions are constructed as spherically symmetric and designed to reproduce the expected properties of
Fornax \cite{Walker_2011, Amorisco_2013, Alexandra_2021}.
The mass models used are based on the Dehnen density profile (see \cref{eq:dehnen_density}) for both the stellar and the halo components.
Also, here, for the system to fully relax before introducing an external potential, the system is run in isolation for $10\times\textrm{ dynamical time}$ for each value of $\beta$.


The initial conditions are given as the best-fit Dehnen parameters of the relaxed profiles with each component mass (see \cref{tab:fornax_parameters_init}).

\begin{table}[ht]
\centering
\begin{tabular}{lccccccc}
\hline \hline $\qquad \beta$ & $0$ & $0.2$ & $0.4$ & $0.6$ & $1.0$ \\
\hline  $a_{\textrm{DM}}/\mathrm{kpc}$ & $5.1$ & $4.0$ & $3.6$ & $3.3$ & $2.7$ \\
 $\gamma_{\textrm{DM}}$ & $1$& $1$ & $1$ & $1$ & $1$ \\
 $a_{\star}/\mathrm{kpc}$ & $0.4$ & $0.3$ & $0.3$ & $0.3$ & $0.3$ \\
 $\gamma_{\star}$ & $1$ & $1$ & $1$ & $1$ & $1$\\
 $M_{\textrm{DM}}(\infty)/{10^{9}}{M_ \odot }$ & $5.4$  & $5.4$ & $5.43$ & $5.4$ & $5.4$ \\
 $M_{\textrm{DM}}\left(r_{\textrm{t}}\right)/{10^{9}}{M_ \odot }$ & $4.0$  & $4.1$ & $4.2$ & $4.6$ & $4.7$ \\
 $M_{\star}(\infty)/{10^{7}}{M_ \odot }$ & $2.4$& $2.4$ & $2.4$ & $2.4$ & $2.4$ \\
 $M_{\star}\left(r_{\textrm{t}}\right)/{10^{7}}{M_ \odot }$ & $2.4$ & $2.4$ & $2.4$ & $2.4$ & $2.4$ \\
\hline \hline
\end{tabular}
\caption{The total and tidally bounded mass with the best-fit Dehnen parameters of the relaxed profiles of Fornax for different $\beta$ values.}
\label{tab:fornax_parameters_init}
\end{table}

\subsection{Orbital Initial Conditions}

In order to set up the initial orbital parameters of Fornax, we performed a backward-time integration from the observed location and velocity of Fornax in order to estimate the appropriate initial conditions for each simulation individually.

The orbit of Fornax within the MW gravitational potential is determined by its current Galactocentric position and velocity, which are inferred from its sky position, radial velocity, distance, and proper motion presented in \cref{tab:current_observation}.
All relevant Fornax parameters (after the relaxation process) are presented in \cref{tab:fornax_parameters_init}.

\begin{table*}[ht]
\centering
\scalebox{0.83}{
    \begin{tabular}{@{}ccccccc@{}}
    \toprule
    \multirow{2}{*}{\textbf{Observation}} &
      {$\alpha$} &
      {$\delta$} &
      {\begin{tabular}[c]{@{}c@{}}distance\\ ($\mathrm{kpc}$)\end{tabular}} &
      {\begin{tabular}[c]{@{}c@{}}${\mu _{{\alpha ^ * }}} $\\ ($\mathrm{mas}  \mathrm{{~yr}^{-1}}$)\end{tabular}} &
      {\begin{tabular}[c]{@{}c@{}}${\mu _\delta }$\\ ($\mathrm{mas}  \mathrm{{~yr}^{-1}}$)\end{tabular}} &
      {\begin{tabular}[c]{@{}c@{}}$v_r$\\ ($\mathrm{km}  \mathrm{{~s}^{-1}}$)\end{tabular}} \\ \cmidrule(l){2-7} 
    & ${2^h}{39^m}{59.3^s}$ & $ - {34^ \circ }{26^{'}}{57^{''}}$ & $147 \pm 12.0$ & $0.374 \pm 0.035$ & $-0.401 \pm 0.035$ & $55.3 \pm 0.3$ \\ \toprule
    \multirow{2}{*}{{\begin{tabular}[c]{@{}c@{}} %
    \textbf{Galactocentric} \\ \textbf{6D coordinates}  \end{tabular}}} &
      {\begin{tabular}[c]{@{}c@{}}X\\ ($\mathrm{kpc}$)\end{tabular}} &
      {\begin{tabular}[c]{@{}c@{}}Y\\ ($\mathrm{kpc}$)\end{tabular}} &
      {\begin{tabular}[c]{@{}c@{}}Z\\ ($\mathrm{kpc}$)\end{tabular}} &
      {\begin{tabular}[c]{@{}c@{}}${v_X}$\\ ($\mathrm{km}  \mathrm{{~s}^{-1}}$)\end{tabular}} &
      {\begin{tabular}[c]{@{}c@{}}${v_Y}$\\ ($\mathrm{km}  \mathrm{{~s}^{-1}}$)\end{tabular}} &
      {\begin{tabular}[c]{@{}c@{}}${v_Z}$\\ ($\mathrm{km}  \mathrm{{~s}^{-1}}$)\end{tabular}} \\ \cmidrule(l){2-7} 
     & -41.2     & -50.9     & -134       & 40.4          & -125           & 82.8    
    \end{tabular}
}
\caption{Observed phase space coordinate   of Fornax\cite{McConnachie_2012,Pietrzyński_2009,Fritz_2018,Schönrich_2010}.}
\label{tab:current_observation}
\end{table*}

\subsection{Simulation Code}

The code
 we develop to integrate and evolve the system in the presence of a fifth force is a hybrid of two \texttt{GADGET2} versions.
The first is \texttt{NGRAV} \cite{CROKER_2016}, and it allows us to determine which forces act on different types of particles, 
Star particles move solely under normal Newtonian gravity, 
while the motion of the DM particles is additionally influenced by the fifth-force computed according to \cref{eq:fifthgrav}.
The second is  \texttt{Gadget2-static potential} \cite{Gadget2-static_potential}, which is used for setting a static potential that affects all particles\footnote{The  code is publicly available at \href{https://github.com/EyalHub-source/Gadget-2-modified-gravity-with-static-potential.git}{\texttt{Modified GADGET2}}.}.
This tool allows us to simulate the potential of the disk and bulge of MW without adding unnecessary particles to the simulation itself.

The simulation ran with a total of $N=10^6$ particles, with $13.8\%$ of them making the Fornax DM halo, $10.6\%$ the stellar component, and $75.6\%$ the MW DM halo.
The forces between particles are smoothed with a Plummer-equivalent softening length of $\epsilon=40\, \textrm{pc}$.

\section{\texorpdfstring{$N$}{N}-body Results}
\label{sec:results}

We have evolved our Fornax-MW $N$-body system using the modified N-body code, starting from the initial conditions described previously in the last section.
This has been done for multiple values of the parameter $\beta$, including the gravity-only scenario with $\beta=0$.

\begin{table}[ht]
\centering
\begin{tabular}{lccccccc}
\hline \hline $\qquad \beta$ & $0$ & $0.2$ & $0.4$ & $0.6$ & $1.0$ \\
\hline
 $M_{\textrm{DM}}\left(r_{\textrm{t}}\right)/{10^{9}}{M_ \odot }$ & $1.4$ & $1.2$ & $1.2$ & $1.2$ & $1.2$\\
 $M_{\star}\left(r_{\textrm{t}}\right)/{10^{7}}{M_ \odot }$ & $2.3$ & $1.7$ & $0.9$ & $0.1$ & \\
  $\Sigma_{\star\textrm{,center}}/{10^{7}}{M_ \odot} \mathrm{kpc}^{-2}$ & $0.8$ & $0.7$ & $0.4$ & $0.1$ &   \\
\hline \hline
\end{tabular}
\caption{Simulation parameters for different $\beta$ values at the last passage near the current estimated location of Fornax.}
\label{tab:simulation_parameters_2}
\end{table}

\begin{figure}[htbp]
\centering
\hspace*{-.8cm}\includegraphics[width=0.5\textwidth]{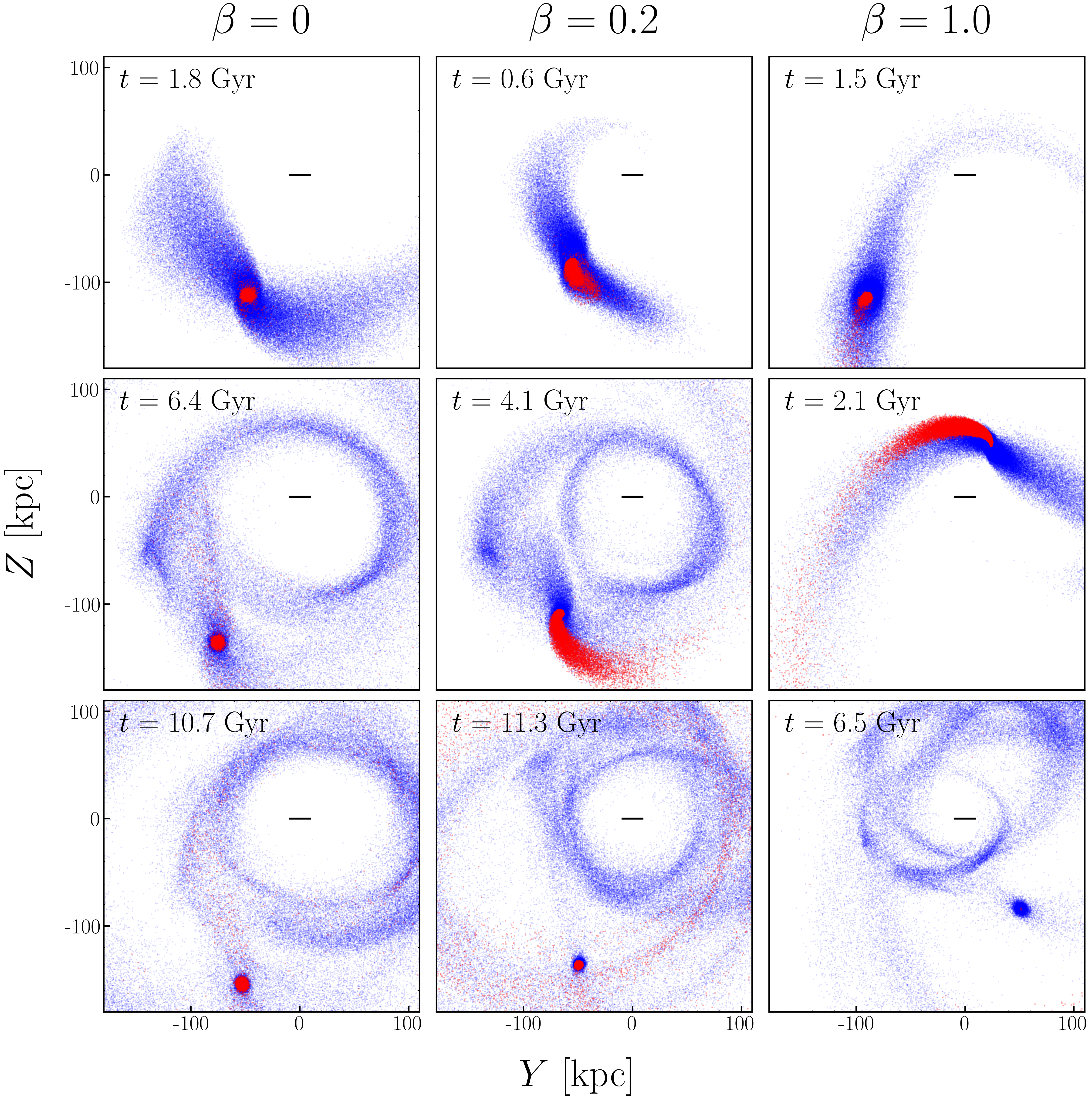}
\caption{Snapshots projected onto the  YZ-Galactic plane illustrating the evolution of the Fornax model under pure gravity ($\beta=0$, left column) and two scenarios with fifth force amplitudes $\beta=0.2$ (middle column) and $\beta=1.0$ (right column). Stellar and DM particles are represented by red and blue dots, respectively. For clarity, only a random subset of particles is plotted.}
\label{fig:morphology}
\end{figure}

We begin with a general assessment of the evolution by means of snapshots of simulation outputs at different times. \Cref{fig:morphology} shows distributions of Fornax stars and DM particles projected onto the YZ-plane at three distinct epochs for simulations with $\beta=0$, $0.2$, and $1$, as indicated.

\begin{figure}[htbp]
    \centering
    \hspace*{-1.0cm}\includegraphics[width=0.52\textwidth]{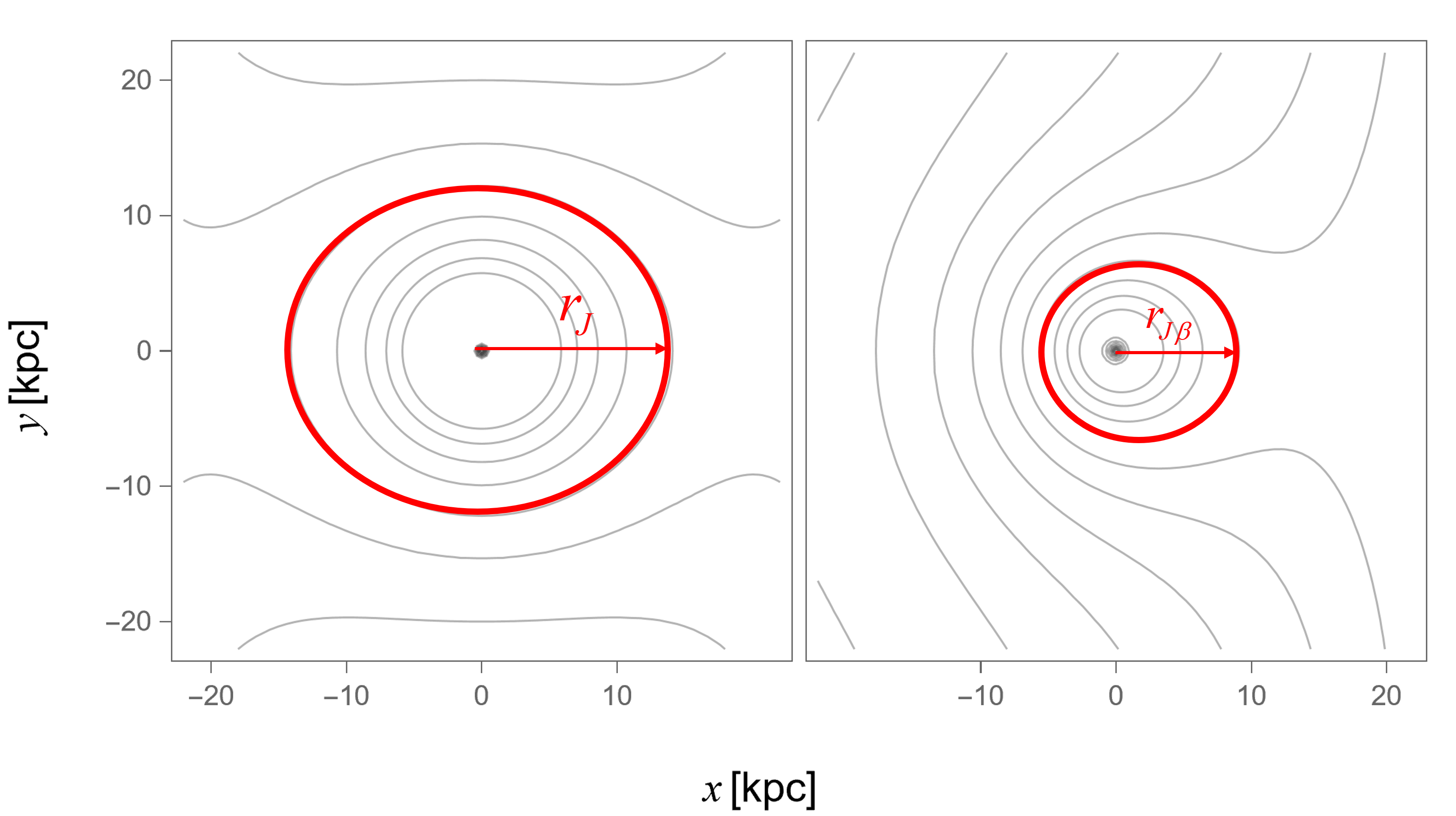}
    \caption{
Contours of the effective potential $\Phi_{\textrm{eff}}(\mathbf{x})$  obtained from the analytic calculations of the MW-Fornax simplified model described in \cref{sec:analyticpotential}. The red contours indicate the last closed zero-velocity surface, where $r_{\rm J}$ and $r_{{\rm J}\beta}$ are the maximum distance from the center of the satellite's mass within this surface. Results are shown for $\beta=0.35$, with the gravitational potential acting on DM (left panel) and on stellar particles (right panel).}
    \label{fig:contours-of-equal-potential}
\end{figure}

In the gravity-only simulation (left column), symmetric tidal streams form in both the DM (blue dots) and stellar (red) components.
This symmetry arises from the structure of the combined gravitational potential, which is illustrated by the equipotential contour plot shown in the left panel of \cref{fig:contours-of-equal-potential}.

The stellar component, being initially more concentrated, undergoes less tidal disruption from the MW gravitational field, resulting in a less extended stellar stream compared to that of the DM.
Introducing a fifth force acting between DM particles substantially alters the dynamics of the stellar component, while leaving the morphology of the DM distribution largely unchanged.
The symmetry of the DM stream is preserved because, from the DM perspective, the fifth force is equivalent to a modification of Newton's gravitational constant.

In contrast, the stellar stream becomes visibly asymmetric at $\beta=0.2$ (middle column), as the DM halo exerts a differential pull on the stars.
This asymmetry is also reflected in the equipotential contours shown in the right panel of \cref{fig:contours-of-equal-potential}.
At this value of $\beta$, the stellar stream disperses significantly after several orbits, as clearly visible at $t=11.3\mathrm{~Gyr}$.
At $\beta=1$, the effect is even more dramatic, leading to the rapid and complete disruption of the stellar component.

The tidal stripping of stars intensifies as the parameter $\beta$ 
 increases\footnote{Consistent with predictions from the first-order solution for ${{r_{\textrm{J}\beta}}}/{r_{\textrm{J}}}$ as given by 
 \cref{eq:small-beta-solution}.}. At every stage of the evolution of the
 satellite, there exists a critical value, $\beta_{\textrm{crit}}$. If 
 $\beta$ exceeds this critical value for a period longer than the dynamical time inside the satellite, then the stellar component will be entirely stripped 
 away, resulting in a complete separation of the halo and stellar 
 components. 
 
For $\beta=1$ at $t=6.5\mathrm{~Gyr}$, the absence of red points in the corresponding panel in \cref{fig:morphology} indicates that the stars have completely escaped the plotted region. This behavior can be interpreted using the virial theorem. 
While moving with the DM halo, the kinetic energy term, $v^2/2$, of the stellar as well as the DM component is approximately half the (fifth+gravity)  potential energy per unit 
mass, which corresponds to $(1+\beta)$ times the gravitational potential
energy per unit mass in the MW potential. For $\beta \gtrsim 1$, this implies that stars are removed with sufficient kinetic energy to exceed their binding energy in the MW 
gravitational field.

\begin{figure}
    \centering
     \hspace*{-.8cm}\includegraphics[width=0.5\textwidth]{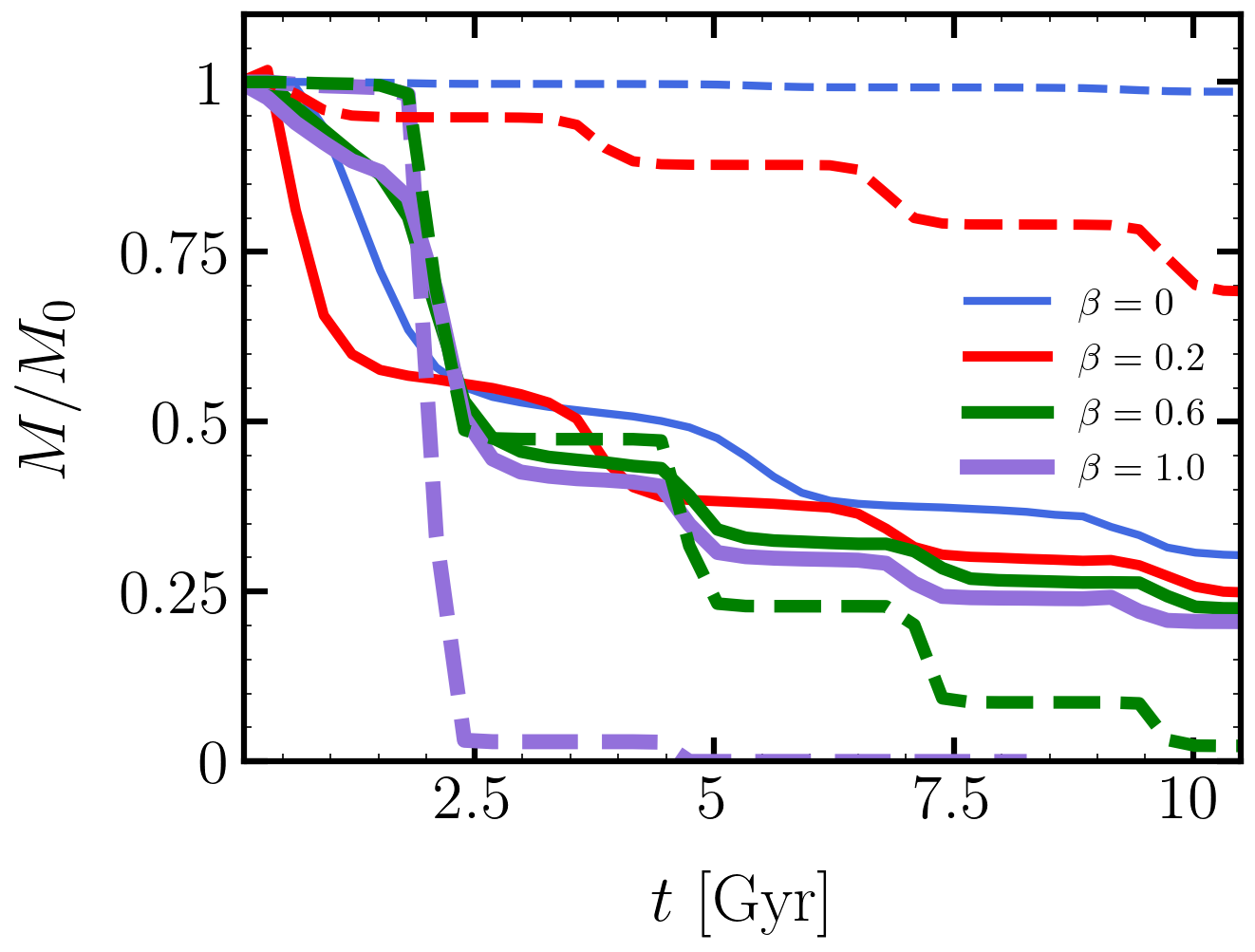}
    \caption{Time evolution of the DM (solid line) and stellar mass (dashed line) enclosed within the tidal radius $r_\textrm{t}$, each normalized to their initial values. 
The  radius $r_\textrm{t}$ is approximated 
using  the Jacobi radius (see \cref{eq:classical-Jacobi-radius}).}
    \label{fig:mass_over_time}
\end{figure}

\Cref{fig:mass_over_time} offers a quantitative assessment of the mass loss as a function of time.
For all plotted values of $\beta$, the solid curves representing $M(t)$ of the DM drop to about 25\% 
of their corresponding initial value by $t \approx 10\mathrm{~Gyr}$. The stripped fraction is independent of $\beta$, and the cumulative mass outside $r_\textrm{t}$ shows nearly identical evolution for all $\beta$ values. In contrast, the dashed curves 
representing the stellar $M(t)$ show a strong dependence on $\beta$.
The stellar mass curves show that tidal effects on the stars become more pronounced as $\beta$ increases; in fact, for $\beta=0$ (dashed blue curve), the stellar mass curve remains flat, indicating negligible mass loss in the absence of the fifth force.

The difference relative to the DM 
arises from the more concentrated initial stellar density profile. Including a fifth force with $\beta = 0.2$ 
results in only a mild decline in the stellar mass, despite the clearly asymmetric stellar stream seen in 
\cref{fig:morphology} for this $\beta$.

The stellar mass loss shown in \cref{fig:mass_over_time} for $\beta = 0.6$ (dashed green) and $\beta = 1$ (dashed purple) becomes substantial by $t \approx 10\mathrm{~Gyr}$, with the mass in both cases falling to nearly zero. For $\beta = 1$, the stellar mass drops sharply to almost zero shortly after $t \approx 2\mathrm{~Gyr}$, following a single pericenter passage. In contrast, for $\beta = 0.6$, the mass decreases in multiple discrete steps associated with successive pericenter passages: roughly 50\% of the stellar mass remains bound after the first passage, but this remnant is gradually stripped over subsequent passages, leaving less than 3\% by the end of the simulation.
Therefore, the cumulative stellar mass converges to progressively lower values as $\beta$ increases.

\begin{figure*}[htbp]
    \centering
    \hspace*{-0.7cm}\includegraphics[width=0.85\paperwidth]{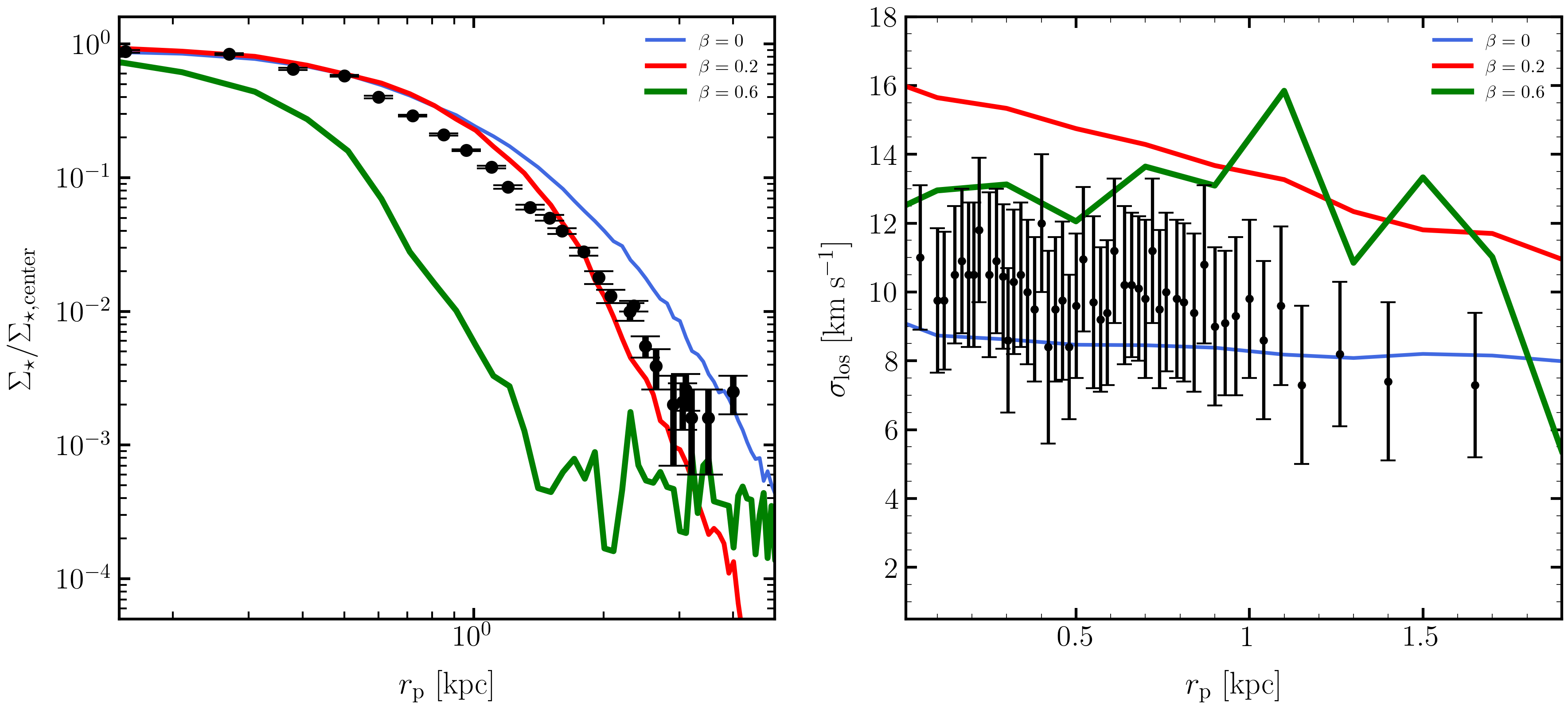}
    \caption{Profiles of stellar surface density (left) and line-of-sight velocity dispersion (right) as functions of projected distance from the satellite center,  for different  $\beta$. Observational data are plotted in black, with surface densities from \cite{Coleman_2005} and velocity dispersions from \cite{Walker_2007}. The surface density profiles are normalized to their central values.}
    \label{fig:profile}
\end{figure*}

\Cref{fig:profile} shows the present-day spherically averaged profiles obtained from our simulations. The halo center is determined using the shrinking-spheres method.
As $\beta$ increases, the surface brightness profile exhibits a sharper radial decline. This trend appears only in simulations where the satellite still retains part of its stellar component. Runs without stars are therefore absent from the left panel.
The observed surface brightness is more closely reproduced by simulations with low $\beta$ values. Moreover, a flat stellar distribution within the satellite core is preserved only for $\beta \lesssim 0.6$, whereas the DM halo retains such a structure for all values of $\beta$ until the end of the evolution.

Complementary insights come from the stellar velocity dispersion profiles in the right panel. The simulated profiles change with $\beta$, reflecting changes in the DM mass distribution.
For $\beta=0$, the velocity dispersion profile remains flat and reproduces the observed mean line-of-sight value of Fornax \cite{Walker_2009}. At higher $\beta$, the agreement with observations becomes only partial, though the overall trends remain consistent with the expected influence of the fifth force.

\section{Discussion \& Conclusions}
\label{sec:conclusion}

Using $N$-body simulations in combination with an analytic model, we have studied the evolution of the Fornax dSph under the influence of a dark-sector fifth force within the Galactic potential.
Our primary goal is to constrain the relative strength $\beta$ of this force compared to gravity by comparing simulated satellites with observations.
The analysis incorporates Fornax’s sky position, distance, radial velocity, and proper motions to reconstruct its orbit within an adopted MW potential, and contrasts the simulated surface-brightness and line-of-sight velocity-dispersion profiles with the observed data.

The relatively low orbital eccentricity of Fornax simplifies the assessment of the long-term impact of the Galactic field on the internal kinematics.
Across all values of $\beta$ considered, the orbit remains only mildly eccentric, with a pericenter of order $\sim 90\, \mathrm{kpc}$ for the adopted MW model.
Fornax is also among the more distant MW dSphs.
These facts make our exclusion of $\beta \gtrsim 0.4$ (not presented within the figures) particularly stringent: closer-in satellites experience stronger tides and would likely display comparable signatures at even smaller $\beta$.

That said, deriving precise constraints from a single system is inherently challenging.
A detailed match between simulations and data is limited not only by observational uncertainties but also by modeling systematics—most notably the MW potential, the orbital reconstruction, and the poorly known assembly history of the MW and its subhalos.

Our simulations nonetheless reveal distinct, testable signatures of a fifth force in the tidal evolution of a dSph.
In particular, increasing $\beta$ enhances stellar stripping relative to the DM, reshaping the surface-brightness profile and suppressing the bound stellar mass, while leaving the total bound DM mass comparatively less affected.

As a complementary check, we developed an analytic treatment of tidal stripping in the presence of a fifth force.
The model neglects several physical effects (dynamical friction, baryonic components of both the satellite and the Galaxy, and non-circular orbits), but it reproduces the qualitative trends and provides intuition for the dominant scalings.
Using the DM distribution from the $\beta=0.2$ simulation as input, the analytic estimate yields a critical $\beta_{\rm crit}\simeq 0.3$ above which stars and DM are expected to segregate (see \cref{Estimation}), consistent with the absence of full segregation in that run.
Applying the same procedure to the DM distribution at $t=4\, \mathrm{Gyr}$ from the $\beta=1$ simulation gives $\beta_{\rm crit}\approx 0.9$, below the simulated value and in agreement with the full segregation seen there.

In systems without a dark-sector fifth force, stellar tidal streams are expected to be approximately symmetric between leading and trailing arms near the progenitor.
A fifth force that violates the equivalence principle induces asymmetries, allowing the symmetry of streams (e.g., in Sgr) to constrain $\beta$ at small values \cite{Kesden_2006oct}.
At large $\beta$, the stellar component can decouple from the DM halo; if a self-bound stellar core survives, its stream can still appear symmetric \cite{Keselman_2009}.
Fornax, however, lacks a detectable stellar stream, so this particular diagnostic is unavailable.
Our constraints, therefore, rely primarily on matching the internal photometric and kinematic observables of Fornax within the modeling assumptions stated above.

We have considered only WEP breaking in this paper. 
A broad class of models involving violation of Lorentz
invariance can also be constrained using cosmological observations 
\citep[e.g.,][]{blas2012}. The dynamics of MW satellites
provide an additional avenue for assessing Lorentz violation, as argued by \citep{Bettoni2017}. These authors showed that the internal 
dynamics of satellites such as Fornax and Draco are sensitive to the presence of a Lorentz-violating field. However, their numerical 
scheme was restricted to a fixed DM distribution. A more detailed modeling, along the lines pursued here, would therefore be
worthwhile.

\begin{acknowledgments}
This research was supported by the Israel Science Foundation (ISF), Grant No. 893/22, and by a grant from the Asher Space Research Institute. 

\end{acknowledgments}

\appendix
\label{appendix:somesort}
\section{Tidal Stripping in the Presence of a Fifth Force: Analytic Considerations}
\label{sec:analyticpotential}

Consider a satellite galaxy moving inside a spherically symmetric host galaxy. We assume that the satellite moves in a circular orbit at a distance $R_{0}$ from the center of the host, where the mass of the satellite is negligible compared to the host's mass within $R_0$.
 In the presence of a fifth force of the form \cref{eq:fifthgrav}, the orbital angular speed of the circular orbit of a DM dominated satellite is,
\begin{equation}
\Omega= \Omega_{0} \sqrt{1+\beta}\; ,
\end{equation}
where
$
\Omega_{0}=\sqrt{{GM}/{R_{0}^{3}}}
$
is the angular frequency in the absence of a fifth force, and we have assumed that the baryonic component is subdominant gravitationally.
 
We adopt a reference frame centered on the satellite center of mass, with the $x$–$y$ plane in  the orbital plane. The unit vector $\hat{\mathbf{e}}{x}$ points toward the host center, and $\hat{\mathbf{e}}{y}$ follows the orbital motion. The frame rotates with angular frequency $\boldsymbol{\Omega} \equiv \Omega \hat{\mathbf{e}}_{z}$, so the acceleration of a stellar or DM particle in the satellite is,
\begin{equation}
\frac{\mathrm{d}^{2} \mathbf{x}}{\mathrm{d} t^{2}}=-\mathbf{f}-2 \boldsymbol{\Omega} \times \frac{\mathrm{d} \mathbf{x}}{\mathrm{d} t}-\boldsymbol{\Omega} \times\left(\boldsymbol{\Omega} \times \mathbf{x}\right),
\label{eq:equations-of-motion}
\end{equation}
where $\mathbf{f}$ is the sum of the forces (per unit mass) resulting from the host halo and the 
satellite itself. We focus on the stripping of stars and henceforth consider only the equations of motion for stellar particles,
\begin{equation}
\mathbf{f} = -\nabla \Phi_{\textrm{s}} - \sum_{k=1}^{3} \Phi_{j k} x_{k} + \beta f_\textrm{DM} \Phi_{j}\, ,
\end{equation}
where $f_\textrm{DM}=M_\textrm{DM}/(M_\star+M_\textrm{DM})$ is the dark matter fraction of the satellite mass. Here $\Phi_{\textrm{s}}(\mathbf{x})$ is the satellite potential, and the second term represents the host potential $\Phi(R)$ expanded to second order around $R_0$. 

In our coordinate system, the center of the host is located at $\mathbf{X}=\left(-R_{0}, 0,0\right)$ and the equations of motion (see \cref{eq:equations-of-motion}) take the form

\begin{align}
&\begin{aligned}\ddot{x}=2 \Omega \dot{y}+\left(\Omega^{2}- \Phi^{\prime \prime}\left(R_{0}\right)\right) x &  \\+  \beta f_\textrm{DM}\Phi^{\prime}\left(R_{0}\right)&- \frac{\partial \Phi_{\textrm{s}}}{\partial x}, \end{aligned} \\
&\ddot{y}=-2 \Omega \dot{x}+\left[\Omega^{2}-\frac{\Phi^{\prime}\left(R_{0}\right)}{R_{0}}\right] y-\frac{\partial \Phi_{\textrm{s}}}{\partial y},  \\
&\ddot{z}=-\frac{\Phi^{\prime}\left(R_{0}\right)}{R_{0}} z-\frac{\partial \Phi_{\textrm{s}}}{\partial z}.
\end{align}

Using the relation $\Phi^{\prime}\left(R_{0}\right)=R_{0} \Omega_{0}^{2}$ we can rewrite $\Omega_{0}^{2}-\Phi^{\prime \prime}\left(R_{0}\right)$ as $-2 R_{0} \Omega_{0} \Omega_{0}^{\prime}\left(R_{0}\right)$ and this in turn can be rewritten as $4 \Omega_{0} A_{0}$ where $A_{0}=$ $A\left(R_{0}\right)$ which is known as the Oort constant and is given by,
\begin{equation}
A(R) \equiv \frac{1}{2}\left(\frac{v_{\textrm{c}}}{R}-\frac{\mathrm{d} v_{\textrm{c}}}{\mathrm{d} R}\right)=-\frac{1}{2} R \frac{\mathrm{d} \Omega_{0}}{\mathrm{d} R},
\end{equation}
thus
\begin{align}
&\begin{aligned} \ddot{x}=2 \Omega_{0} \sqrt{1+\beta} \dot{y}+\left(\beta \Omega_{0}^{2}+4 \Omega_{0} A_{0}\right)& x +\\ \beta f_\textrm{DM}R_{0}\Omega_{0}^{2}&-\frac{\partial \Phi_{\textrm{s}}}{\partial x}, \end{aligned} \\
& \ddot{y}=-2 \Omega_{0} \sqrt{1+\beta} \dot{x}+\beta \Omega_{0}^{2} y-\frac{\partial \Phi_{\textrm{s}}}{\partial y},  \\
& \ddot{z}=-\Omega_{0}^{2} z-\frac{\partial \Phi_{\textrm{s}}}{\partial z}\, .
\end{align}

The motion is associated with the followng  integral of motion,
\begin{equation}
    E =\frac{1}{2} v^{2}+\Phi_{\textrm{eff}}(\mathbf{x})\, .
\end{equation}
The form of $\Phi_{\textrm{eff}}$ has different forms for DM and stellar particles since 
 the DM feel the additional force derived from $\Phi_{\textrm {scalar }}=\beta \Phi_{\textrm {halo }}$. For the stellar particles we have,
\begin{multline}
    \Phi_{\textrm{eff}}(\mathbf{x}) = -2 \Omega_{0} A_{0} x^{2}-\frac{1}{2}\Omega_{0}^{2} \left(\beta x^{2}+\beta y^{2}-z^{2}\right) \\ -\beta f_\textrm{DM}R_{0}\Omega_{0}^{2}x+\Phi_{\textrm{s}}(\mathbf{x})\, ,
    \label{eq:stellar-effective-potential}
\end{multline}
while  for the DM, 
\begin{equation}
    \Phi_{\textrm{eff}}(\mathbf{x}) = (1+\beta) \left(\frac{1}{2}\Omega_{0}^{2} z^{2}-2 \Omega_{0} A_{0} x^{2}+\Phi_{\textrm{s}}(\mathbf{x})\right)\, .
    \label{eq:DM-effective-potential}
\end{equation}

For $\beta \neq 0$, the inclusion of the additional force modifies the stellar zero-velocity surface, defined by $\Phi_{\textrm{eff}}(\mathbf{x})=E$. In contrast, the DM distribution follows the non-additional force model, differing only by a factor of $(1+\beta)$ in the integral of motion $E$ (see \cref{fig:contours-of-equal-potential}).

\section{Tidal Stripping of stars}

Assume the satellite potential $\Phi_{\textrm{s}}$ arises from a spherical mass distribution $M_{\textrm{s}}(r)$ centered at $\mathbf{x}=0$, and the host is also spherical with enclosed mass $M(R)$ at radius $R$.
We rewrite the expression for 
the acceleration of 
a  DM test particle as (for $y=\dot{y}=z=\dot{z}=0$),
\begin{equation}
    \ddot x = \left( {\beta {\Omega _0}^2 + 4{\Omega _0}{A_0}} \right)x + \frac{{G{M_{\textrm{s}}}}}{{{{\left| x \right|}^3}}}x\left( {p - 1} \right) + \beta {R_0}{\Omega _0}^2\, ,
    \label{eq:The-stationary-points-equation}
\end{equation}
where 
    $p\equiv{\mathrm{d}\ln M_{\textrm{s}}}/{\mathrm{~d}\ln r}$.
Taking  $\ddot x=0$,  \cref{eq:The-stationary-points-equation} yields  analogs to the Lagrange points $L_{2}$ and $L_{3}$ in the restricted three-body problem.
For  $\beta =0$ the solutions reduce to $x=\pm x_{\textrm{J}}$, where 

\begin{equation}
\label{eq:xjnew}
    x_{\textrm{J}}=  \left(\frac{G M_{\textrm{s}}}{4 \Omega_{0} A_{0}}\right)^{1 / 3}{{(1 - p)^{1/3}}}\; .
\end{equation}

For   $\Phi_{\textrm{s}}$ and $\Phi(R) $ corresponding  point masses $m$ and $M$ respectively, \cref{eq:xjnew} reduces to  usual expression for the Jacobi radius,
$x_{\textrm{J}}= r_{\textrm{J}}$ where 
\begin{equation}
r_{\mathrm{J}} \equiv \left(\frac{G m}{4 \Omega_{0} A_{0}}\right)^{1 / 3}=\left(\frac{m}{3 M}\right)^{1 / 3} R_{0}\, .
    \label{eq:classical-Jacobi-radius}
\end{equation}

We will now explore how  \cref{eq:The-stationary-points-equation} is influenced by the additional force acting on the  DM only.
We will consider now a general mass distribution of the satellite in the form of,
\begin{equation}
    {M_{\textrm{s}}}\left( r \right) = m\:f\left( r \right)\, .
\end{equation}

The Jacobi radius of an orbiting stellar system corresponds to the observed tidal radius, i.e. the maximum extent of the satellite. To determine this, we consider the zero-velocity surface of the last closed contour enclosing the satellite center, obtained by solving the equation for $x>0$ (\cref{fig:contours-of-equal-potential}). For simplicity, we assume the satellite is dominated by dark matter ($f_\textrm{DM} \approx 1$, as in Fornax), which allows us to write \cref{eq:The-stationary-points-equation} in the dimensionless form,
\begin{equation}
\frac{{{d^2}\chi }}{{d{\tau ^2}}} = \Gamma \xi  + (\xi  + 1)\chi  + \left( {p\left( {\left| \chi  \right|} \right) - 1} \right)f\left( {\left| \chi  \right|} \right)\frac{{\left| \chi  \right|}}{{{\chi ^3}}}\, ,
\label{eq:dimensionless_stationary-points}
\end{equation}
where $\tau  \equiv 2\sqrt {{\Omega _0}{A_0}}t ,\:\xi  \equiv \beta {\Omega _0}^2{r_{\textrm{J}}}^3/({G{M_\textrm{s}}}) = \beta {{\Omega _0}}/({4{A_0}})$, 
${\rm{  }}\chi  \equiv x/{{r_{\textrm{J}}}},\:  {\rm{  }}\Gamma  \equiv {{R_0}}/{{r_{\textrm{J}}}}$. It is convenient to write \cref{eq:dimensionless_stationary-points} as,
\begin{equation}
\begin{aligned}
\frac{d\chi}{d\tau} &= u \, , \\
\frac{du}{d\tau} &= \Gamma \xi + (\xi + 1)\chi + (p - 1) f \frac{|\chi|}{\chi^3} \, .
\end{aligned}
\label{eq:phase-plane}
\end{equation}
These equations represent the phase plane in terms of $\chi$ and $u$.
We identify the point with $d u / d \tau =0$ as the ratio ${{r_{\textrm{J}\beta}}}/{r_{\textrm{J}}}$. This ratio is useful since satellite systems generally do not follow circular orbits 
and the standard derivation of the Jacobi radius cannot be directly applied. Nonetheless,  the tidal radius can still be approximated using the Jacobi radius (\cref{eq:classical-Jacobi-radius}), with $R_{0}$ replaced by the pericenter distance \citep[e.g.,][]{King_1962}. Under this assumption, the ratio ${{r_{\textrm{J}\beta}}}/{r_{\textrm{J}}}$ remains unchanged for non-circular orbits, since no parameter in \cref{eq:small-beta-solution} is particularly sensitive to this adjustment.

For sufficiently low $\beta$ we expect $r_{\textrm{J}\beta}\approx r_\textrm{J}$.
In this limit, linearizing the last equations  near $\chi  = 1$ and identified $r_{\textrm{J}\beta}$ as the point with $d u/d\tau=0$ we find,  
 \begin{equation}
\frac{{{r_{\textrm{J}\beta}}}}{{{r_{\textrm{J}}}}} \cong 1 - \frac{{\left( {3  - \gamma} \right)a/r_{\textrm{J}}  + \left( {\Gamma  + 1} \right)\xi }}{{3 + 3\left( {\Gamma  + 1} \right)\xi }} \, .
\label{eq:small-beta-solution}
 \end{equation}
This expression decays with increasing $\beta $, however, it is reliable only for $\beta \lesssim 0.1$.

\smallskip

\section{Estimation of Critical \texorpdfstring{$\beta$}{beta}}
\label{Estimation}
We use \cref{eq:small-beta-solution} to estimate the
stellar tidal radius for $\beta \ll 1$, and determine the critical value $\beta = \beta_\textrm{crit}$ at which the DM and stellar components become completely separated. The answer follows from analyzing the phase
plane 
defined by \cref{eq:phase-plane}. This system
exhibits a fixed point ($u = d u / d \tau = 0$), $\chi^*$, 
near the origin, whose stability determines 
$\beta_\textrm{crit}$. Linearization around the fixed point 
yields a traceless Jacobian matrix, with stability governed 
by its eigenvalues. For $\beta < \beta_\textrm{crit}$, the 
eigenvalues are purely imaginary, and the fixed point 
behaves as a stable center. For $\beta > 
\beta_\textrm{crit}$, the eigenvalues are real with opposite 
signs, yielding a saddle point. This transition is 
characterized by the determinant $\Delta$,
\begin{equation}
\Delta \left( \chi  \right) =\! -\! 1\! -\! \xi (\beta)\!  -\! \frac{{\left| \chi  \right|}}{{{\chi ^3}}}f  p\left( {p\! -\! 3\! +\! \frac{2}{p}\! +\! \frac{{d\ln p}}{{d\ln \left| \chi  \right|}}} \right)\, ,
\label{eq:Jacobian_det}
\end{equation}
where $\beta_\textrm{crit}$ is fixed by the condition $\Delta(\chi^*) = 0$, thus we obtain
\begin{equation}
    {\beta _\textrm{crit}}\! =\!   \frac{{4{A_0}}}{{{\Omega _0}}}\!\left[ { \frac{{\left| {\chi^ * } \right|}}{{{\chi^*}^3}}fp\!\left( {3\! -\! p\! -\! \frac{2}{p}\! -\! {{ {\frac{{d\ln p}}{{d\ln \left| \chi  \right|}}} \Bigg|}_{\chi^* }}}\! \right)}\!-\!1\! \right]{\mkern 1mu}\, .
\end{equation}

\bibliographystyle{apsrev4-2}
\bibliography{Bibliography}

\end{document}